\documentclass[conference]{IEEEtran}
\IEEEoverridecommandlockouts

\usepackage{amsmath,amssymb,amsfonts}
\usepackage{algorithmic}
\usepackage{graphicx}
\usepackage{textcomp}
\usepackage{booktabs}
\usepackage{hyperref}
\usepackage{xcolor}
\usepackage{multirow}
\usepackage{array}
\usepackage{dsfont}
\usepackage{cite}
\usepackage{pifont}
\usepackage{icomma}
\hypersetup{
    colorlinks=true,
    linkcolor=blue,
    filecolor=magenta,      
    urlcolor=cyan,
    pdfpagemode=FullScreen,
    }

\def\BibTeX{{\rm B\kern-.05em{\sc i\kern-.025em b}\kern-.08em
    T\kern-.1667em\lower.7ex\hbox{E}\kern-.125emX}}
\begin{document}

\title{Speaker-IPL: Unsupervised Learning of Speaker Characteristics with i-Vector based Pseudo-Labels}

\author{
\IEEEauthorblockN{\begin{tabular}{c}Zakaria Aldeneh$^{\dagger}$, Takuya Higuchi$^{\dagger}$, Jee-weon Jung$^{\ddagger}$, Li-Wei Chen$^{\ddagger}$, Stephen Shum$^{\dagger}$,\\ Ahmed Hussen Abdelaziz$^{\dagger}$, Shinji Watanabe$^{\ddagger}$, Tatiana Likhomanenko$^{\dagger}$, Barry-John Theobald$^{\dagger}$\end{tabular}}
\IEEEauthorblockA{\textit{$^{\dagger}$ Apple} \qquad \textit{$^{\ddagger}$Carnegie Mellon University}
\\
\{zaldeneh, takuya\_higuchi\}@apple.com}}

\maketitle
\begin{abstract}
\textit{Iterative self-training}, or \textit{iterative pseudo-labeling} (IPL)---using an improved model from the current iteration to provide pseudo-labels for the next iteration---has proven to be a powerful approach to enhance the quality of speaker representations. Recent applications of IPL in unsupervised speaker recognition start with representations extracted from very elaborate self-supervised methods (e.g., DINO). However, training such strong self-supervised models is not straightforward (they require hyper-parameter tuning and may not generalize to out-of-domain data) and, moreover, may not be needed at all. To this end, we show that the simple, well-studied, and established i-vector generative model is enough to bootstrap the IPL process for the unsupervised learning of speaker representations. We also systematically study the impact of other components on the IPL process, which includes the initial model, the encoder, augmentations, the number of clusters, and the clustering algorithm. Remarkably, we find that even with a simple and significantly weaker initial model like i-vector, IPL can still achieve speaker verification performance that rivals state-of-the-art methods.

\end{abstract}

\section{Introduction}
\label{sec:intro}
Speech representations that capture unique speaker characteristics are useful for many tasks, including speaker verification, diarization, speech enhancement, text-to-speech, and more~\cite{jung2024espnet}. State-of-the-art approaches for extracting these representations rely on supervised models---models trained to classify speakers from a closed set. While effective, the success of supervised methods hinges upon having access to data that are annotated with accurate speaker identities.
The requirement for annotated data has made self-supervised learning (SSL) an attractive alternative for training speaker models. Consequently, several SSL methods have been proposed to learn speaker representations without the reliance on ground-truth annotations~\cite{heo2022self,fathan2024self,chen2023comprehensive,zhang2022c3,cho2022non,thienpondt2020idlab,chen2024self,tao2022self,zhou2024self}. Representations from these methods can be used directly ``out-of-the-box'', or they can be used to bootstrap pseudo-labels for an \textit{iterative self-training} (or \textit{iterative pseudo-labeling}, IPL) procedure to further refine the representations: an initial model (\textit{teacher}) provides pseudo-labels for training a new better model (\textit{student}), while the latter becomes a \textit{teacher} in the next iteration and the process is repeated until a \textit{student} model stops improving over its \textit{teacher}~\cite{xu2020iterative,hsu2021hubert,chiu2022self}.

Recent applications of IPL in speaker recognition use representations extracted from strong SSL methods (e.g., DINO~\cite{cho2022non}). However, training these SSL methods is very elaborate; they require careful tuning of several hyper-parameters, including  batch size,  augmentations, network architecture, etc., and may not generalize well to out-of-domain data.
The difficulty that comes with training strong SSL speaker models raises the following questions: \textit{do we really need such a strong model? Can we use a ``simple'' unsupervised speaker model to bootstrap the IPL procedure?}
The benefit of using a simple speaker model is that it reduces the need for carefully designing specialized unsupervised models.
Instead, we can use well-studied and established models to bootstrap the iterative training procedure.

In this work, we study the utility of the traditional i-vector~\cite{dehak2010front} approach for bootstrapping the IPL procedure. The i-vector model is a well-studied \textbf{unsupervised technique} for modeling the super-vectors from a Gaussian Mixture Model (GMM) using factor analysis and the Expectation-Maximization (EM) algorithm. The approach provides an alternative to current strong self-supervised neural approaches as it requires optimizing fewer hyper-parameters and is based on well-established statistical models. The contributions of this work are two-fold: (1) we demonstrate that i-vectors can be used in place of state-of-the-art DINO embeddings in the IPL procedure while achieving competitive performance; and (2) we run a systematic analysis into the components that impact the performance of the IPL framework, which includes the initial model, the encoder, augmentations, the number of clusters, and the clustering algorithm. Our results highlight the effectiveness of IPL and provide practitioners with the necessary analysis to understand the impact of performance on various aspects of the framework.

\section{Related Work}
\label{sec:related}
Recent work in the unsupervised speaker verification literature can be broadly grouped into three categories. The first category includes works that proposed and made a case for methods that do not use the IPL framework (e.g.,~\cite{chen2023pushing,zhang2022c3}), citing the requirement for defining the number of clusters apriori and its dependency on the quality of the representations from the initial model as a limitation. Our experiments address both of these concerns about IPL: we study the impact of under- and over-estimating the number of speakers in the clustering stage, as well as the dependency of the IPL framework on the initial model. The second category includes work that acknowledged the utility of IPL but focused solely on improving the initial model (e.g.,~\cite{tu2024contrastive,cho2022non,jung2022pushing,heo2022self}). Our work studies the \textit{necessity} of a powerful initial model and asks if the i-vector model can be used to bootstrap the iterative process. Finally, the third category includes work that focused on refining the quality of the pseudo-labels and improving the clustering method (e.g.,~\cite{tao2022self,zhou2024self,cai2021iterative,han2023self,chen2023unsupervised,thienpondt2020idlab,fathan2024self}). Our work presents a systematic study into the components that impact the performance of the IPL framework, including the initial model, the encoder, augmentations, number of clusters, and clustering algorithm.

\section{Unsupervised Speaker-IPL}
\label{sec:approach}

\begin{figure}[ht]
\begin{minipage}[b]{1.0\linewidth}
  \centering
\def\svgwidth{0.6\columnwidth}
\small
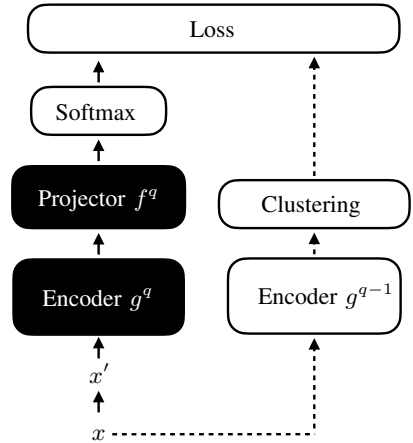
\end{minipage}
\caption{The unsupervised speaker iterative pseudo-labeling (speaker-IPL) framework. For a given iteration $q$, we train a speaker encoder $g^q(\cdot)$ and a projector $f^q(\cdot)$ to predict pseudo-labels generated by clustering representations from the encoder $g^{q-1}(\cdot)$. $x'$ is the \textit{augmented speech segment} and $x$ is the unaltered speech sample. $g^0(\cdot)$ is an unsupervised i-vector model. Blocks in black indicate trainable components, while other blocks indicate non-trainable components.}
\label{fig:framework}
\vspace{-0.3cm}
\end{figure}

Our unsupervised speaker-IPL framework is shown in Fig.~\ref{fig:framework}. At every iteration $q$, we train a speaker encoder $g^q(\cdot)$ and a projector $f^q(\cdot)$ to predict pseudo-labels generated by clustering representations from the previous iteration encoder $g^{q-1}(\cdot)$. \textbf{The representations from $g^{0}(\cdot)$ are obtained from the unsupervised i-vector generative model}. The input to the $g^{q-1}(\cdot)$ encoder is the original unaltered sample $x$, while the input to the encoder $g^q(\cdot)$ is an \textit{augmented segment} $x'$ of the original sample. \textit{The segmentation and augmentation process exploits the speaker stationarity assumption in the input sample}\footnote{Later, we show that segmentation and augmentation improve the model performance though are not so crucial for the speaker-IPL.} (i.e., any segment from the original sample inherits the same speaker label despite variations from phonetic content). The iterative training procedure is repeated multiple times to improve the quality of the extracted representations~\cite{xu2020iterative}. 

The speaker encoder $g^{q}(\cdot)$ maps a variable-length input to a fixed-size representation that captures the characteristics of different speakers. This mapping is achieved through frame-level modeling followed by utterance-level pooling~\cite{jung2024espnet}. Recent neural architectures for modeling frame-level variations use time-delay neural networks, convolutional neural  network, and transformers (e.g., ~\cite{zeinali2019but,desplanques2020ecapa,zhang2022mfa}). Common pooling techniques include statistics pooling and attentive statistics pooling~\cite{okabe2018attentive}.

We use the unsupervised i-vector approach as our initial model $g^{0}(\cdot)$ to provide the first set of pseudo-labels. i-vector decomposes a speaker- and session-dependent super-vector $M$ using the following equation: $$M=m+Tw$$ where $m$ is the speaker- and session-independent super-vector from the universal background model (UBM); $T$ is the rectangular matrix defining the total variability space; and $w$ is a low-dimensional random intermediate vector (i-vector) with prior $\mathcal{N}(0, I)$. The variabilities not captured by $T$ are captured by the covariance matrix $\Sigma$ of the model. We refer the reader to the original i-vector paper~\cite{dehak2010front} for more details.

\section{Experiments}
\label{sec:experiments}
We design our experiments to (1) assess the effectiveness of i-vectors for bootstrapping IPL and (2) study the various components that impact the performance of IPL. We compare the results we obtain to results from prior works.

\subsection{Setup}
\label{sec:setup}
\subsubsection{Data} 
We use the VoxCeleb$2$~\cite{chung2018voxceleb2} development set, containing $1,092,009$ utterances from $5,994$ speakers, to train our unsupervised methods without speaker labels and Vox$1$-O, Vox$1$-E, and Vox$1$-H trial lists to evaluate verification performance. We include the VoxSRC-$20$ test trial list as an additional test set.

\subsubsection{IPL} We train the encoder $g^q(\cdot)$ for iteration $q$ following the recipe described in~\cite{zhang2022mfa}. Specifically, we use a batch size of $200$, the ADAM optimizer~\cite{kingma2014adam} with an initial learning rate of $0.008$, a weight decay of $10^{-8}$, linear warm-up for the first $2$k steps, and the additive margin softmax loss with a $0.2$ margin and a scaling factor of $30$. The number of epochs is set to $20$. After training concludes, we pick the encoder from the epoch that achieved the best validation performance on Vox$1$-O and cluster its features to provide pseudo-labels for the next iteration. 

\subsubsection{Encoder $g^q (\cdot)$}
We evaluate two state-of-the-art encoders in our experiments: MFA-Conformer~\cite{zhang2022mfa} and ECAPA-TDNN~\cite{desplanques2020ecapa}. Both encoders take in $80$-dimensional Mel-spectrogram features extracted with a window length of $25$ms and a $10$ms hop size. We set the embedding size to $192$ for the two models and use the same hyper-parameters as~\cite{zhang2022mfa}. The MFA-Conformer and ECAPA-TDNN have $21.2$M and $22.2$M trainable parameters, respectively.

\begin{table}[t]
  \centering
  \caption{i-vectors are effective for bootstrapping pseudo-labels for training an unsupervised speaker verification model using IPL. The equal error rates (EERs, \%) are reported on the Vox1-O.}
  \label{tab_results_1}
  \begin{tabular}{lr}
    \toprule
    \textbf{Models}    & \textbf{EER (\%)}\\
    \midrule
    Fully-supervised (upper bound) & $0.82$\\
    DINO &  $4.53$\\
    i-vector (lower bound)&  $13.95$\\
    \midrule
    DINO $+$ CL~\cite{heo2022self} &  $4.47$\\
    CAMSAT~\cite{fathan2024self} &  $3.06$\\
    DINO $+$ aug.~\cite{chen2023comprehensive} &  $2.51$\\
    C$3$-DINO$_2$~\cite{zhang2022c3} &  $2.50$\\
    R-DINO~\cite{chen2023pushing} &  $3.29$\\
    SDPN w/ DR~\cite{chen2024self} &  $1.80$\\
    DINO ($3$ iters.) ~\cite{cho2022non} &  $2.13$ \\
    MoCo ($7$ iters.)~\cite{thienpondt2020idlab} &  $2.10$ \\
    SCL ($2$ iters.)~\cite{tao2022self} &  $1.66$\\
    DINO + AT + HT ($5$ iters.)~\cite{zhou2024self} &  $1.35$\\
    \midrule
    i-vector + IPL ($11$ iters.)\\
    \quad$\rightarrow$ w/ ECAPA-TDNN    & $1.79$\\
    \quad$\rightarrow$ w/ MFA-Conformer & $1.14$\\
    \bottomrule
  \end{tabular}
  \vspace{-0.2cm}
\end{table}

\subsubsection{Augmentations $x \rightarrow x'$} Our augmentation strategy during speaker-IPL includes additive noise and reverberation. For additive noise, we use samples from the MUSAN corpus~\cite{snyder2015musan} and mix them with SNRs sampled from $\text{Uniform}(10\text{dB}, 25\text{dB})$; for reverberation, we use simulated room impulse responses (RIRs) from~\cite{ko2017study}.

\subsubsection{Clustering} We follow a clustering approach similar to that described in~\cite{cho2022non}, and use \texttt{faiss}~\cite{douze2024faiss} to learn $25$k centroids with $k$-means and then use \texttt{scikit-learn}~\cite{scikit-learn} to learn $7.5$k centroids using agglomerative hierarchical clustering (AHC) with average linkage and cosine similarity metric~\cite{douze2024faiss,scikit-learn}.

\subsubsection{i-vector}
We follow the standard Kaldi recipe to train the i-vector extractor.\footnote{https://github.com/kaldi-asr/kaldi/tree/master/egs/voxceleb/v1.} Specifically, we train the UBM with $2048$ components and full covariance matrices on $72$-dimensional Mel-frequency cepstral coefficients, which include delta and double-delta coefficients, and then train a $400$-dimensional i-vector extractor using the longest $100$k utterances from the training set. Unlike the Kaldi recipe, we only use samples from the VoxCeleb2 development set for training our i-vector extractor. We length-normalize the i-vectors before using them in our framework~\cite{garciaromero11_interspeech}. Note that no augmentations or speaker labels are used when training the i-vector model.

\subsubsection{DINO}\label{DINO_setup} We also train a DINO model to compare against i-vectors. The DINO model uses the MFA-Conformer encoder and is learned using an ADAM optimizer with an initial learning rate of $0.0025$. Two long ($4$ seconds) and four short ($2$ seconds) segments are randomly cropped and used for teacher and student models, respectively, during training. We use noise and RIR augmentation with SNRs randomly sampled from $\text{Uniform}(5\text{dB}, 20\text{dB})$. The temperature parameters were set as $0.04$ and $0.1$ for the teacher and the student, respectively. The momentum gradually increases from $0.996$ to $1$ via the cosine scheduler. Following~\cite{chen2023comprehensive}, the projection head is three fully-connected layers with a hidden size of $2048$ and a bottleneck dimension of $256$ followed by a fully-connected layer mapping the bottleneck dimension to the final output dimension $K=8000$. In contrast to i-vector training, DINO training requires tuning more hyper-parameters and is very sensitive to their values.

\begin{figure}[t]
\begin{minipage}[b]{0.85\linewidth}
  \centering
\def\svgwidth{\linewidth}
\footnotesize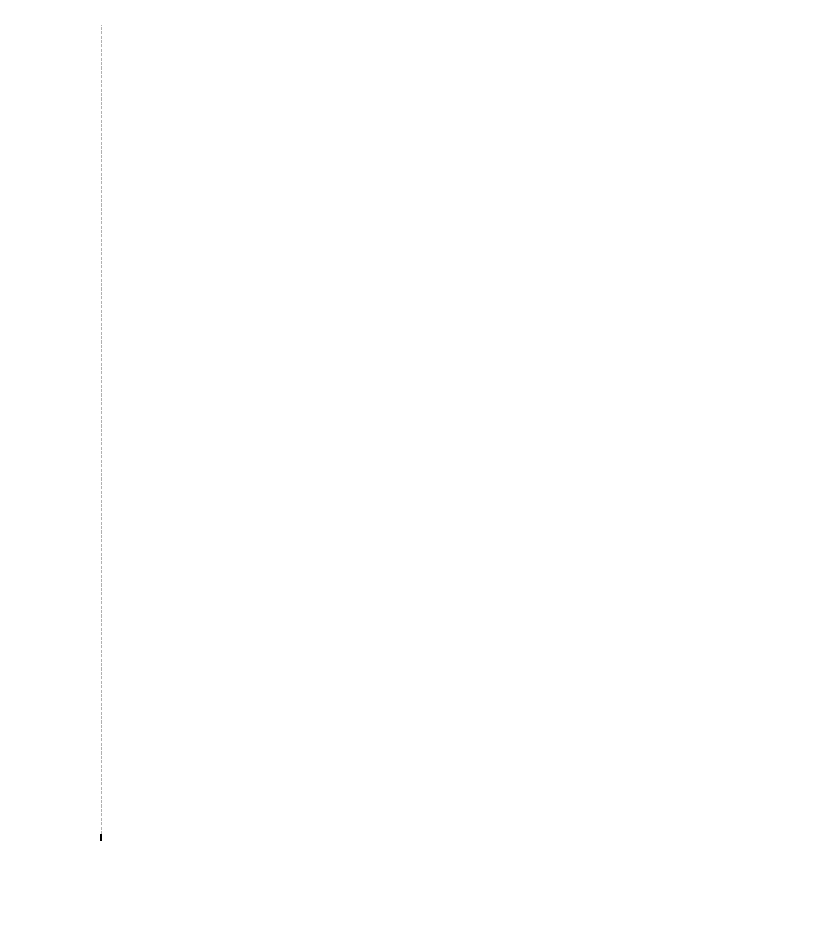
\end{minipage}
\caption{Changing components of the speaker-IPL impacts both performance and convergence trends. The performance is reported on the Vox$1$-O.}\label{fig:eer_vs_clusters}
\end{figure}

\begin{figure}[h]
\begin{minipage}[b]{0.85\linewidth}
  \centering
\def\svgwidth{\linewidth}
\footnotesize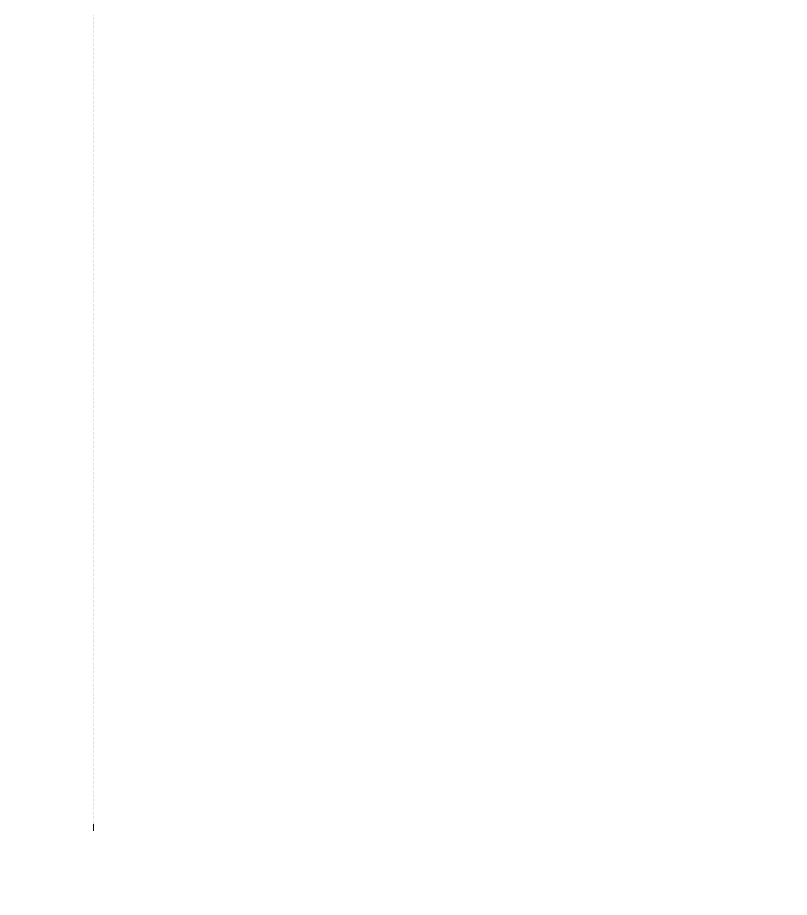
\end{minipage}
\caption{Changing components of the speaker-IPL impacts both performance and convergence trends. The performance is reported on the VoxSRC-$20$ (test).}\label{fig:eer_vs_clusters2}
\vspace{-0.2cm}
\end{figure}

\subsection{Speaker-IPL: Bootstrapping IPL with i-Vectors}\label{subsec:pseudo}
The first experiment aims to study the effectiveness of the i-vector model for bootstrapping IPL and compare its performance to what is reported in prior work. The results are summarized in Table~\ref{tab_results_1}. 

\begin{table*}[t]
  \centering
  \caption{Evaluating the iteration that gave the lowest equal error rate (EER, $\%$) on Vox$1$-O for each setup from Fig.~\ref{fig:eer_vs_clusters}.}
  \label{compare_dino}
  \resizebox{\linewidth}{!}{%
  \begin{tabular}{llccccccccc}
    \toprule
     \multirow{2}{*}{\textbf{}}& \multirow{2}{*}{\textbf{Init. Model}} & \multirow{2}{*}{\textbf{Encoder}} & \multirow{2}{*}{\textbf{Clustering}} & \multirow{2}{*}{\textbf{Aug.}}  & \multirow{2}{*}{\textbf{\# Clusters}}     & \multicolumn{4}{c}{\textbf{EER (\%)}}\\
     & & & & & & Vox$1$-O & Vox$1$-E & Vox$1$-H & VoxSRC-$20$\\
    \midrule
     A.$1$ & i-vector & MFA-Conformer & AHC & \checkmark & $7.5$k & $1.06$ & $2.16$ & $3.61$ & $7.05$\\
     
     A.$2$ & i-vector & MFA-Conformer & AHC & \checkmark & $3$k & $1.99$ & $3.71$ & $6.37$ & $9.34$\\
     A.$3$ & i-vector & MFA-Conformer & AHC & \checkmark & $12$k & $1.70$ & $2.46$ & $4.17$ & $7.80$\\

     A.$4$ & i-vector & MFA-Conformer & AHC & \ding{55} & $7.5$k & $1.31$ & $2.71$ & $4.37$ & $8.33$\\
     A.$5$ & i-vector & ECAPA-TDNN    & AHC & \checkmark & $7.5$k & $1.78$ & $2.89$ & $4.98$ & $8.71$\\
     A.$6$ & i-vector & MFA-Conformer & $k$-means & \checkmark & $7.5$k & $2.30$ & $3.42$ & $5.93$ & $9.80$\\
     B.$1$ & DINO & MFA-Conformer & AHC  &  \checkmark &  $7.5$k & $1.12$ & $1.86$ & $3.29$ & $6.92$\\
    \bottomrule
  \end{tabular}}
  \vspace{-0.4cm}
\end{table*}

Our supervised MFA-Conformer baseline establishes the upper bound performance with an equal error rate (EER) of $0.82\%$. We obtain an EER of $4.53\%$ when scoring the features from our DINO implementation with cosine similarity, and an EER of $13.95\%$ when scoring the $400$-dimensional i-vectors with cosine similarity. The i-vector performance establishes the lower bound performance our iterative method uses as a starting point. Bootstrapping the IPL with i-vectors and running it for $11$ iterations gives an EER of $1.79\%$ and $1.14\%$ when using the ECAPA-TDNN and the MFA-Conformer encoders, respectively. With MFA-Conformer the EER is lower than $12/13$ of the methods in Table~\ref{tab_results_1}, and with ECAPA-TDNN, the EER is lower than $10/13$ of the methods.
This result underscores the effectiveness of IPL and shows that it can be used with less powerful initial models and still rival state-of-the-art methods.

Our approach is not the only one from Table~\ref{tab_results_1} that uses IPL---\cite{cho2022non,thienpondt2020idlab,tao2022self,zhou2024self} also do. However, our approach is \textbf{the only one} to show that IPL can achieve state-of-the-art performance even if we start with a less powerful i-vector model. In the next section, we run a systematic study into the components that impact the performance of IPL, including the initial model, the encoder, augmentations, number of clusters, and clustering algorithm.

\subsection{IPL Ablations}\label{subsec:additional}
We start with the i-vector setup from the previous section (setup A.$1$ from Figs.~\ref{fig:eer_vs_clusters}, \ref{fig:eer_vs_clusters2}, and Table~\ref{compare_dino}) and study how changing components in the IPL framework impacts the convergence trends and  performance. 

\subsubsection{Adjusting the number of clusters} Setups A.$2$ and A.$3$ in Figs.~\ref{fig:eer_vs_clusters} and \ref{fig:eer_vs_clusters2} show how adjusting the number of learned i-vector clusters impacts EERs. Our base setup (A.$1$) uses $7.5$k clusters (similar to~\cite{cho2022non})  and we evaluate two additional setups, one with $12$k clusters and another with $3$k
clusters. We obtain EERs of $2.24\%$, $1.14\%$, and $1.7\%$ on Vox$1$-O after running IPL for $11$ iterations with $3$k, $7.5$k, and $12$k clusters, respectively. Note that the ground-truth number of speakers for the VoxCeleb$2$ development set is $5,994$, suggesting that \textbf{over-estimating the number of speakers has a less detrimental effect on performance than under-estimating them.}

\subsubsection{Removing the augmentations} Not applying augmentations when training the model (setup A.$4$) raises the EERs by $23.58\%$ and $18.16\%$ on Vox$1$-O and VoxSRC-$20$, respectively. Without augmentations, the model needs an additional iteration to converge to the best performance on Vox$1$-O. Note that augmentations are only used when training the IPL framework, and they are not used when training the initial i-vector model. ~\cite{cho2022non} showed that removing augmentations from DINO raised the EER by $381\%$ on Vox$1$-O, suggesting that \textbf{our framework is less sensitive to augmentation.}

\subsubsection{Changing the encoder}
Replacing the MFA-Conformer with ECAPA-TDNN (setup A.$5$) raises the EERs by $67.92\%$, $33.80\%$, $37.95\%$, and $23.55\%$ for the Vox$1$-O, Vox$1$-E, Vox$1$-H, and VoxSRC-$20$ trial lists, respectively. This result shows\textbf{ that the choice of encoder is important for IPL}. The framework gives an EER of $1.06\%$ and $1.78\%$ on Vox$1$-O after six iterations with ECAPA-TDNN and eight iterations with MFA-Conformer, respectively. 

\subsubsection{Changing the clustering algorithm} Using $k$-means in setup A.$6$ to cluster the representations from each step into $7.5$k clusters directly without using AHC raises the EERs by $116.98\%$, $58.33\%$, $64.27\%$, and $39.01\%$ for the Vox$1$-O, Vox$1$-E, Vox$1$-H, and VoxSRC-$20$ trial lists. This result demonstrates that \textbf{the choice of clustering algorithm has the highest impact on performance when keeping everything else fixed.}

\subsubsection{Using a stronger initial model} Finally, we replace the i-vector model with a more powerful DINO model (setup B.$1$) that we trained as described in Section~\ref{sec:setup}. The DINO model gives an EER of $4.53\%$ when scoring its features directly on Vox$1$-O; a noticeable improvement over $13.95\%$ obtained from the raw i-vector model. 
Although strong initialization with DINO achieves better performance in several test sets, the performance difference is still marginal, or weak initialization with i-vector achieved better performance on Vox1-O.
So, we conclude that \textbf{the weak initialization with i-vector can rival the strong initialization}. Also, considering the elaborate nature of the initial model based on DINO (see Section~\ref{DINO_setup}), the proposed methods have substantial benefits in practice.

\section{Conclusion}
This paper presented a systematic study of iterative pseudo-labeling for unsupervised learning of speaker representations. We studied the impact of the initial model, the encoder, augmentations, and clustering algorithm on the overall performance. We showed that the iterative pseudo-labeling procedure can achieve competitive performance compared to prior work with the less powerful i-vector initial model while having a similar convergence rate. Also, other hyper-parameters (e.g., clustering, encoder type) are more important for unsupervised learning of speaker representations than the choice of initial model.


\bibliographystyle{IEEEtran}
\bibliography{refs}

\end{document}